\input phyzzx.tex
\tolerance=1000
\voffset=-0.0cm
\hoffset=0.7cm
\sequentialequations
\def\rl{\rightline}

\def\t1{{\tilde 1}}

\def\t{\theta}

\def\S{Schwarzschild~}

\REF{\BEK}{J. Bekenstein, Lett. Nuov. Cimento {\bf 4} (1972) 737; Phys Rev. {\bf D7} (1973) 2333; Phys. Rev. {\bf D9} (1974) 3292.}
\REF{\HAW}{S. Hawking, Nature {\bf 248} (1974) 30; Comm. Math. Phys. {\bf 43} (1975) 199.}
\REF{\EUC}{G. Gibbons and S. H. Hawking, Phys. Rev. {\bf D15} (1977) 2752.}
\REF{\CON}{L. Susskind, [arXiv:hep-th/9309145].}
\REF{\WAL}{R. M. Wald, Phys. Rev. {\bf D48} (1993) 3427, [arXiv:gr-gc/9307038]; V. Iyer and R. M. Wald, Phys. Rev. {\bf D50} (1994) 846, [arXiv:gr-qc/9403028]; Phys. Rev. {\bf D52} (1995) 4430, [arXiv:gr-qc/9503052].}
\REF{\SBH}{E. Halyo, A. Rajaraman and L. Susskind, Phys. Lett. {\bf B392} (1997) 319, [arXiv:hep-th/9605112].}
\REF{\EDI}{E. Halyo, Int. Journ. Mod. Phys. {\bf A14} (1999) 3831, [arXiv:hep-th/9610068]; Mod. Phys. Lett. {\bf A13} (1998), [arXiv:hep-th/9611175].}
\REF{\DES}{E. Halyo, [arXiv:hep-th/0107169]; JHEP {\bf 0112} (2001) 005, [arXiv:hep-th/0108167]; [arXiv:hep-th/0308166].}
\REF{\EDIH}{E. Halyo, [arXiv:1403.2333]; [arXiv:1406.5763].}
\REF{\CARL}{S. Carlip, Phys. Rev. Lett. {\bf 82} (1999) 2828, [arXiv:hep-th.9812013]; Class. Quant. Grav. {\bf 16} (1999) 3327,
[arXiv:gr-qc/9906126].}
\REF{\SOL}{S. Solodukhin, Phys. Lett. {\bf B454} (1999) 213, [arXiv:hep-th/9812056].}
\REF{\LAST}{E. Halyo, [arXiv:1502.01979], [arXiv:1503.07808]; [arXiv:1506.05016]; [arXiv:1606.00792].}
\REF{\HOL}{G. 't Hooft, [arXiv:gr-qc/9310026]; L. Susskind, J. Math. Phys. {\bf 36} (1995) 6377, [arXiv:hep-th/9409089]; R. Bousso, Rev. Mod. Phys. {\bf 74} (2002) 825, [arXiv:hep-th/0203101].}
\REF{\ADS}{J. Maldacena, Adv. Theor. Math. Phys. {\bf 2} (1998) 231, [arXiv:hep-th/9711200]; S. Gubser, I. Klebanov and A. Polyakov, Phys. Lett. {\bf B428} (1998) 105, [arXiv:hep-th/9802109]; E. Witten, Adv. Theor. Math. Phys. {\bf 2} (1998) 253, [arXiv:hep-th/9802150].}
\REF{\YAR}{O. Aharony, S. S. Gubser, J. Maldacena, H. Ooguri and Y. Oz, Phys. Rep. {\bf 323} (2000) 183, [arXiv:hep-th/9905111].}
\REF{\NHO}{H. K. Kunduri, J. Lucetti and H. S. Reall, Class. Quant. Grav. {\bf 24} (2007) 4169, [arXiv:0705.4214];
P. Figueras, H. K. Kunduri, J. Lucetti and M. Rangamani, Phys. Rev. {\bf D78} (2008) 044042, [arXiv:0803.2998].}
\REF{\DAB}{A. Dabholkar, A. Sen and S. P. Trivedi, JHEP {\bf 0701} (2007) 096, [arXiv:hep-th/0611143].}
\REF{\TAK}{T. Azeyanagi, T. Nishioka and T. Takayanagi, Phys.Rev. {\bf D77} (2008) 064005, [arXiv:0710.2956].}
\REF{\ENT}{S. Ryu and T. Takayanagi, Phys. Rev. Lett. {\bf 96} (2006) 181602, [arXiv:hep-th/0603001; JHEP {\bf08} (2006) 045,
[arXiv:hep-th/0605073]; T. Nishioka, S. Ryu and T. Takayanagi, JPhys. {\bf A42} (2009) 504008, [arXiv:0905.0932].}
\REF{\CQM}{D. Gaiotto, A. Strominger and X. Yin, JHEP {\bf 0511} (2005) 017, [arXiv:hep-th/0412322].}
\REF{\CHI}{A. Strominger, JHEP {\bf 9901} (1999) 007, arXiv:[hep-th/9809027]; V. Balasubramanian, J. de Boer,  M. M. Sheikh--Jabbari and J. Simon, JHEP {\bf 1002} (2010) 017, [arXiv:0906.3272];}
\REF{\SEN}{R. K. Gupta and A. Sen, JHEP {\bf 0904} (2009) 034, [arXiv:0806.0053]; Entropy {\bf 13} (2011) 1305, [arXiv:1101.4254].}
\REF{\SOL}{S. Solodukhin, Living Rev.Rel. {\bf 14} (2011) 8, [arXiv:1104.3712] and references therein.}
\REF{\REN}{L. Susskind and J. Uglum, Phys. Rev. {\bf D50} (1994) 2700, [arXiv:hep-th/9401070].}
\REF{\JAC}{T. Jacabson, [arXiv:gr-qc/9404039].}
\REF{\RMW}{R. M. Wald, General Relativity (1983), The Chicago University Press.}
\REF{\IRA}{J. Sadeghi and V. R. Shajiee, [arXiv:1512.01817].}
\REF{\CAD}{M. Cadoni, Mod.Phys.Lett. {\bf A21} (2006) 1879, [arXiv:hep-th/0511103].}
\REF{\CAR}{J. L. Cardy, Nucl. Phys. {\bf B463} (1986) 435.}
\REF{\REV}{M. Rangamani and T. Takayanagi, [arXiv:1609.01287].}
\REF{\DOB}{A. Lewkowycz and J. Maldacena, JHEP {\bf1308} (2013) 090, [arXiv:1304.4926].}
\REF{\GB}{T. Clunan, S. F. Ross and D. J. Smith, Class. Quant. Grav. {\bf 21} (2004) 3447, [arXiv:gr-qc/0402044].}
\REF{\ORD}{X. Dong, JHEP {\bf 1401} (2014) 044, [arXiv:1310.5713]; J. Camps, JHEP {\bf 1403} (2014) 070, [arXiv:1310.6659].}
\REF{\LAS}{E. Halyo, in preparation.}
\REF{\EPR}{J. Maldacena and L. Susskind, Fortsch. Phys. {\bf 61} (2013) 781, [arXiv:1306.0533].}
\REF{\KIN}{W. Kinnersley, J. Math. Phys. {\bf 14} (1973) 651; S. Bertini, S. L. Cacciatori and D. Klemm, Phys. Rev. {\bf D85} (2012)
064018, [arXiv:1106.0999].}

\singlespace
\rl{SU-ITP-17/08}
\pagenumber=0
\normalspace
\medskip
\bigskip
\titlestyle{\bf{The Holographic Entanglement Entropy of Schwarzschild Black Holes}}
\smallskip
\author{ Edi Halyo{\footnote*{e--mail address: halyo@stanford.edu}}}
\smallskip
\centerline {Department of Physics} 
\centerline{Stanford University} 
\centerline {Stanford, CA 94305}
\smallskip
\vskip 2 cm
\titlestyle{\bf ABSTRACT}

We show that \S black hole metrics are, asymptotically, Weyl equivalent to $AdS_2 \times S^{D-2}$ where the spherical factor is the horizon. The holographic entanglement entropy of $AdS_2$ exactly reproduces the \S black hole entropy which implies that black hole degrees of freedom live at asymptotic infinity. In generalized theories of gravity, the same procedure reproduces Wald entropy.

\singlespace
\vskip 0.5cm
\endpage
\normalspace

\centerline{\bf 1. Introduction}
\medskip

The origin of black hole entropy, especially those of \S black holes[\BEK,\HAW], remains one of the biggest puzzles in gravity.  There are numerous methods for computing black hole entropy such as those which use Euclidean gravity[\EUC], the conical deficit angle[\CON], the Noether charge[\WAL], the dimensionless Rindler energy[\CON,\SBH-\EDIH], horizon CFTs among others[\CARL-\LAST]. In all these methods, one assumes that the black hole degrees of freedom are located at or near the horizon. This certainly makes sense in the context of holography[\HOL] which postulates that a black hole is described by degrees of freedom on its boundary, i.e. the horizon. In addition, all these methods exploit the fact that the near horizon geometry is Rindler space.

On the other hand, there is an alternative notion of holography according to which the fundamental degrees of freedom of a black hole live on a screen at infinity. The celebrated AdS/CFT correspondence[\ADS,\YAR] which is the only fully--fledged realization of holography is a concrete example. In the AdS/CFT correspondence, the degrees of freedom of an AdS black
hole are located not on its horizon but on the AdS boundary (where the finite temperature boundary CFT at lives). 
At first thought, it is difficult to see how this notion of holography can apply to \S black holes since their metrics
are asymptotically flat. Thus, at infinity, there is no boundary. Moreover, flat space does not carry any degrees of freedom that can account for black hole entropy.

Quite separately, it is well--known that the near horizon geometries of all extremal black holes have either an $AdS_2$ or $AdS_3$
(which is an $S^1$ fibration of $AdS_2$)[\NHO] component which is the origin of their entropies[\DAB]. Again, it is
difficult to see how this can be relevant for \S black holes since their near horizon geometry is Rindler space. A hopeful sign is the fact that Rindler space and $AdS_2$ are Weyl equivalent after a coordinate transformation. Since the near horizon region of Rindler space and the boundary of $AdS_2$, exactly the regions where entropy originates, are described by CFTs one may hope to relate \S black hole entropy to $AdS_2$ by a Weyl transformation.

In this paper, we argue that even though \S black hole metrics are asymptotically flat, they can be Weyl transformed to  space--times which are asymptotically $AdS_2 \times S^{D-2}$ where $S^{D-2}$ is in the original horizon directions and $R_{AdS_2}=R_{S^{D-2}}=r_0$ with $r_0$ the \S radius. We show that the temperature and entropy of both \S black holes and the asymptotically $AdS_2 \times S^{D-2}$ space--times are given by the same functions of the black hole mass. Thus,
we assume that these two space--times have the same thermodynamics or are at least in the same universality class.
At low energies (compared to $1/R_{S^{D-2}}$) we can integrate out the spherical directions and find that the asymptotic physics is described by $AdS_2$. It is well--known that global $AdS_2$ has two disconnected boundaries and is described by a pure state of the boundary theory in which the degrees of freedom on both boundaries are entangled[\TAK]. Therefore, if we restrict ourselves
to only one boundary and trace over the other one, we obtain a mixed state with a nonzero entanglement entropy. This has been computed
by the holographic entanglement entropy method[\ENT] which gives the entanglement entropy of a region $A$ on the boundary as
$$S_{ent}(A)= {Area(\Sigma_A) \over {4G}} \quad, \eqno(1)$$
where $Area(\Sigma_A)$ is the area of the codimension two minimal surface in the bulk such that the boundaries of $A$ and $\Sigma_A$ coincide. In our case, the boundary of $AdS_2$ is one dimensional and therefore the minimal surface is a point in the bulk with 
$Area(\Sigma_A)=1$.  Therefore, using eq. (1) we get[\TAK]
$$S_{ent}(AdS_2)={1 \over {4 G_2}}= {A_{H} \over {4 G_D}} \quad, \eqno(2)$$
where we used $G_2=G_D/A_{H}$ and the fact that the horizon and $S^{D-2}$ have the same area, $A_H$. 
Thus, the entanglement entropy of $AdS_2$ precisely reproduces the entropy of the \S black hole. This result can be obtained by a more precise calculation of the holographic entanglement entropy as we show in the next section. This method can also be applied
to black holes in generalized theories of gravity and correctly reproduces the Wald entropy[\TAK]. 
The two holographic computations of $S_{ent}$ that we describe in the next section are technically very similar to the calculations of black hole entropy using the conical deficit angle[\CON] and Euclidean gravity[\EUC] methods even
though conceptually they are very different.

Our results indicate that the degrees of freedom of \S black holes reside at asymptotic infinity just like those of AdS black holes.
In this case, the screen at infinity seems to be an $S^{D-2}$ at the boundary of $AdS_2$ space--time.
This is only true for the Weyl transformed \S metric but as we mentioned above these two space--times have the same thermodynamics and therefore we assume that the origin of their entropies is also the same. The $AdS_2$ boundary theory has been described by conformal quantum mechanics[\CQM], chiral or light--cone 2D CFTs[\CHI] and strings that live on $AdS_2$[\SEN]. Nevertheless, we do not have a clear understanding of the black hole degrees of freedom that live at the boundaries of $AdS_2$.
It is important to note that our method requires only that the original black hole metric be asymptotically flat. Thus, we can easily generalize our results to all asymptotically flat nonextremal black objects.  

The above method should not be conflated with the description of black hole entropy due to the entanglement of modes just inside and outside of the horizon which has a rich literature[\SOL]. As mentioned above, in our case the degrees of freedom are at asymptotic infinity. Moreover, the conventional entanglement entropy is a UV cutoff dependent quantity that diverges in the continuum limit. Our result is finite and gives the correct entropy. It seems that holographic entanglement entropy has a natural cutoff in $AdS_2$ that
leads to the correct entropy. Conventional entanglement entropy is actually a (one--loop) correction to black hole entropy[\REN] and
reproduces the total black hole entropy only in the context of induced gravity[\JAC]. In our case, gravity is not induced. 

This paper is organized as follows. In section 2, we show that the entropy of \S black holes is given by that
of an $AdS_2$ space--time which is at the asymptotic infinity of the Weyl transformed \S metric. In section 3, we generalize
our results to theories of gravity beyond General Relativity and show that our method reproduces the Wald entropy for black holes.
In section 4, we discuss the calculational similarities and conceptual differences between our method and those of conical deficit angle and Euclidean gravity. Section 5 includes a discussion of our results and our conclusions.

\bigskip
\centerline{\bf 2. The Holographic Entanglement Entropy of Schwarzschild Black Holes}
\medskip

In this section, we show that the asymptotic infinity of a D--dimensional \S black hole metric is Weyl equivalent to
$AdS_2 \times S^{D-2}$. At low energies, i.e. for $E <<1/R_{S^{D-2}}$,
we can integrate out the $S^{D-2}$ and the asymptotic physics is described by the $AdS_2$ factor. It is well--known that global $AdS_2$ has two disconnected boundaries and is described by a pure state in which the states of the two boundary theories are entangled[\TAK]. If we are constrained to only one boundary, we have to trace over the states of the second one.
This leads to a mixed state with an entanglement entropy which has been computed in ref. [\TAK]. We show that the $AdS_2$ entanglement entropy exactly reproduces the \S black hole entropy. This result implies that the (unknown) degrees of freedom of the black hole reside at asymptotic infinity and not near the horizon.

Consider a D--dimensional \S black hole with the metric
$$ds^2=-\left(1-{r_0^{D-3} \over r^{D-3}}\right) dt^2+ \left(1-{r_0^{d-3} \over r^{D-3}}\right)^{-1} dr^2+ r^2 d\Omega^2_{D-2} \quad,
\eqno(3)$$
where the radius of the black hole is given by $r_0^{D-3}= 16 \pi G_D M/(D-2) A_{D-2}$. We now Weyl transform this metric by
$$g_{\mu \nu} \to \Omega^{2}(r) g_{\mu \nu}= {r_0^2 \over r^2} g_{\mu \nu} \quad, \eqno(4)$$
followed by the coordinate transformation $r \to r_0^2/r$[\IRA]. The Weyl transformation is chosen to keep the area of $S^{D-2}$
fixed and equal to the horizon area. It turns the metric into a direct product of a $2D$ metric and $S^{D-2}$ of fixed radius. The
coordinate transformation exchanges large ($r>r_0$) and small ($r<r_0$) $r$ keeping the horizon at $r_0$ fixed. The region outside the black hole now corresponds to $r<r_0$ with asymptotic infinity at $r=0$.

Under these two transformations the \S metric becomes
$$ds^2=-\left({r^2 \over r_0^2}-{r^{D-1} \over r_0^{D-1}} \right) dt^2+  \left({r^2 \over r_0^2}-{r^{D-1} \over r_0^{D-1}} \right)^{-1} dr^2 + r_0^2 d\Omega^2_{D-2} \quad. \eqno(5)$$
This is the metric of an $AdS_2$ black hole times $S^{D-2}$ but with an unusual range for $r$, $0 \leq r \leq r_0$[\CAD].
It is well--known that General Relativity is not a Weyl invariant theory[\RMW]. Therefore, Weyl transformed solutions to the Einstein equation do not remain solutions. (They are however solutions to the transformed Einstein equation.) The metric in eq. (5) is a
is a solution only for $D=4$ when it describes an $AdS_2$ black hole times $S^2$. In this case, the metric factorizes into two $2D$ metrics and in two dimensions the Einstein equation is trivially satisfied by all metrics. For $D>4$, 
the solutions describe an $AdS_2$ black hole times $S^{D-2}$ where the $S^{D-2}$ part is no longer a solution (to the $D-2$ dimensional Einstein equation). We will nevertheless continue to use eq. (5) for all $D$ since, as we show below, if we ignore this subtlety we can obtain the entropy of \S black holes for all $D$. 

The metric in eq. (5) has two horizons: the black hole horizon at $r=r_0$ and the $AdS_2$ horizon
at $r=0$ (which corresponds to the asymptotic infinity, $r \to \infty$ in the \S metric). The near horizon region of the black hole is described by Rindler space which can be described by a horizon CFT with central charge $c=12 E_R$ where $E_R$ is the dimensionless Rindler energy[\LAST]. The black hole corresponds to a state with $L_0=E_R$ in the horizon CFT[\LAST] and its entropy is correctly given by the Cardy formula[\CAR].

In this paper, we are interested in describing the black hole in the other, asymptotic limit. 
In the limit $r \to 0$ the metric in eq. (5) becomes (for $D \geq 4$) 
$$ds^2=-{r^2 \over r_0^2} dt^2+ {r_0^2 \over r^2} dr^2 + r_0^2 d\Omega^2_{D-2} \quad. \eqno(6)$$
This is exactly the metric of $AdS_2 \times S^{D-2}$ with the radii $R_{AdS_2}=R_{S^{D-2}}=r_0$. We now claim that the entropy of the \S black hole is given by the entanglement entropy of the $AdS_2 \times S^{D-2}$[\TAK]. This is due to the fact that the
Weyl transformation in eq. (4) (plus the coordinate transformation $r \to r_0^2/r$) does not change either the temperature or entropy of the \S black hole. Thus both space--times in eqs. (3) and (6) have the same thermodynamics.
The horizon area and entropy are not invariant under a generic Weyl transformation. However, under the Weyl transformation in eq. (4), the area of the $S^{D-2}$ is precisely that of the \S
black hole horizon since $\Omega^{2}(r)r^2=r_0^2$ by construction.  
Therefore, the entropies corresponding to the metrics in eqs, (3) and (6) are the same. 
The Weyl transformation in eq. (4) is the simplest one that keeps the horizon entropy invariant and leads to a metric that is a direct product and not a fibration. 

It is also easy to show that the temperature is invariant under any Weyl transformation and in particular under that in eq. (4). This is can be seen from the definition
of Hawking temperature, $T_H=f^{\prime}(r_0)/4 \pi$ where $f(r)$ is the coefficient of the $dt^2$ term in the \S metric. 
Another way to see this is to notice that
the near horizon geometry and therefore the temperature is not modified by a Weyl transformation. 
Consider the generic metric for a nonextremal black hole
$$ds^2=-f(r) dt^2+ f(r)^{-1}dr^2 + r^2 d\Omega^2_{D-2} \quad. \eqno(7)$$
After a Weyl transformation $ds^2 \to \Omega^{2}(r) ds^2$ and taking the near horizon limit $r=r_0+y$ where $y<<r_0$ we find
$f(r)=f^{\prime}(r_0)y$ and $\Omega^{2}(r)=\Omega^{2}(r_0)+ \Omega^{2 \prime} (r_0) y$. To lowest order in $y$ the metric becomes
$$ds^2={{y \Omega^{2}(r_0)} f^{\prime}(r_0)} dt^2+ {\Omega^{2}(r_0) \over {f^{\prime}(r_0)y}} dy^2 + \Omega^{2}(r_0)r^2 d \Omega^2 \quad. \eqno(8)$$
In terms of the proper distance to the horizon defined by $d \rho=\sqrt{\Omega^{2}(r_0)/f^{\prime}(r_0) y} dy$ we find that the near horizon metric is
$$ds^2=-{\rho^2 f^{\prime}(r_0)^2\over 4} dt^2+ d \rho^2+ \Omega^{2}(r_0)r^2 d \Omega^2 \quad. \eqno(9)$$
We see that the Weyl transformation does not modify the $t-\rho$ directions of the metric and thus the Hawking temperature 
is invariant under conformal transformations.  

We have established the \S and the asymptotically $AdS_2 \times S^{D-2}$ metrics in eqs. (3) and (6) have the same entropy. Therefore,
we can try to explain the \S black hole entropy using the physics of $AdS_2 \times S^{D-2}$. At low energies (compared to $1/r_0$) we can integrate out the $S^{D-2}$ degrees of freedom and are left with pure $AdS_2$ with the reduced Newton constant $G_2=G_D/A_H$ where $A_H$ is the common area of the horizon and $S^{D-2}$.

The Weyl transformation in eq. (4) and the inversion of $r$ have been considered in ref. [\IRA] before. However, in that work, the entropy of $AdS_2$ was obtained in the boundary CFT of a dilatonic $AdS_2$ black hole rather than that in eq. (5) which is in pure 2D gravity which is not dynamical. There is no justification for considering an $AdS_2$ black hole in dilatonic gravity when the dimensional reduction results in pure 2D gravity. Moreover, it is strange that, in ref. [\IRA], pure $AdS_2$ is described by an excited state of the boundary CFT (even though it seems to give the right black hole entropy). In this paper, we remain in pure 2D gravity as required by eq. (5) and the entropy of $AdS_2$ vacuum arises from entanglement between its two boundaries.

The $t-r$ part of the metric in eq. (6) can be seen as both the metric of the Poincare patch of $AdS_2$ and that of global $AdS_2$[\YAR] 
$$ds^2=-\left(1+{r^2 \over r_0^2}\right) dt^2+\left(1+{r^2 \over r_0^2}\right)^{-1} dr^2 \quad, \eqno(10)$$
near its boundary, i.e. for $r>>r_0$. We will assume the latter since the Poincare patch has only one boundary and therefore no entanglement entropy. However,
global $AdS_2$, as mentioned above, has an entanglement entropy which may potentially (and indeed does) match those of \S black holes.  
We see that the asymptotic infinity of \S black holes is described by the boundary of global $AdS_2$ (times $S^{D-2}$).
Global $AdS_2$ is also described by the metric
$$ds^2=r_0^2 {{-dt^2+d \sigma^2} \over sin^2\sigma} \quad. \eqno(11)$$
This metric, unlike higher dimensional $AdS$ space--times, has two disconnected (one dimensional) boundaries at $\sigma=0,\pi$.
In the $AdS_2$ vacuum, the degrees of freedom on the two boundaries are entangled[\TAK].
The total Hamiltonian of global $AdS_2$ is given by the sum $H_{tot}=H_1 + H_2$ where $H_{1,2}$ are the (unknown) Hamiltonians that describe the copies of conformal quantum mechanics living on each boundary. The $AdS_2$ vacuum is a pure but entangled state given by
$$|\Psi_{AdS}>= \sum_{i,j} c_i |i>_1 \otimes |i>_2 \quad, \eqno(12)$$
where $|i>_1$ ($|i>_2$) is the eigenstates of $H_1$ ($H_2$).
If we are restricted to only one boundary, then we need to trace over the states of the second one. As a result, the state in eq. (12) becomes a mixed state described by the density matrix 
$$\rho_1=Tr_2 \rho_{tot}=Tr_2 \sum_{i,j} |c_i|^2 (|i>_1 \otimes |i>_2) (<i|_2 \otimes <i|_1) \quad.  \eqno(13)$$ 
The entanglement entropy is then given by[\REV]
$$S_{ent}=-Tr(\rho_1log \rho_1)=-{\partial \over \partial n} log (Tr \rho_1^n)|_{n=1} \quad. \eqno(14)$$
Using the holographic entanglement entropy formula, $S_{ent}$ can be computed as in eq. (2)[\ENT]. However, we can also compute
$Tr \rho_1^n$ directly in the bulk by using the replica trick[\TAK], i.e. using the n--sheeted $AdS_2$ space in the path integral computation. $AdS_2$ is basically a strip with two boundaries at $\sigma=0,\pi$. The n--sheeted $AdS_2$ is obtained by introducing  a cut which when crossed over takes us from one sheet of $AdS_2$ to the next. For entanglement entropy, the region of interest on the boundary is a point and the minimal surface in the bulk is a line that ends in the bulk (since only one boundary is traced over). The easiest way to see this is to start with the entanglement entropy of a region $A$ i.e. an arc on the boundary $S^1$
of $AdS_3$ and compactify it on the $S^1$. Then, $A$ and $A^c$ become the boundaries of $AdS_2$, the minimal curve becomes a point in the bulk and the area inside it becomes a line that connects the bulk to the boundary.
This line is exactly the cut required for the n--sheeted $AdS_2$ which introduces a conical deficit angle of 
$2 \pi (1-n)$. Then, by the AdS/CFT duality, the density matrix (to the $n^{th}$ power) is given by
$$Tr \rho_1^n= e^{-I_{EH}(AdS_{2,n})} \quad, \eqno(15)$$
where the 2D gravity Einstein--Hilbert action $I_{EH}(g)$ is obtained by dimensionally reducing the 4D gravity action over the $S^{D-2}$ with constant radius and $AdS_{2,n}$ is the n--sheeted Euclidean $AdS_2$.
From eq. (15) we find
$$log (Tr \rho_1^n)=-I_{EH}(AdS_{2,n}) \quad, \eqno(16)$$
where $I_{EH}$ computed over the n--sheeted (Euclidean) $AdS_2$ is given by
$$I_{EH}=-{1 \over {16 \pi G_2}} \int_{AdS_{2,n}} d^2x (R+ \Lambda) \quad, \eqno(17)$$
with $G_2=G_D/A_H$. The curvature scalar, including the contribution arising from the conical singularity due to the cut, is 
$$R_{AdS_{2,n}}=R_{AdS_2}+4 \pi (1-n) \delta(x) \quad. \eqno(18)$$
The properly normalized entanglement entropy is
$$S_{ent}=-{\partial \over \partial n} (-I_{EH}+nI_0) \quad, \eqno(19)$$
where $I_{EH}$ is given by eq. (17) and $I_0$ is the Einstein--Hilbert action over one copy of $AdS_2$. Using eqs. (17)-(19) 
we obtain
$$S_{ent}={1 \over {4 G_2}}= {A_H \over 4 G_D}=S_{BH} \quad, \eqno(20)$$
which is exactly the \S black hole entropy.

The above computation of $S_{ent}$ of $AdS_2 \times S^{D-2}$ is quite straightforward but perhaps too slick since it depends only on the dimensionality of space. A more precise and detailed derivation of eq. (20) can be given which is along the lines of the proof of the holographic entanglement entropy formula[\REV,\DOB]. We begin by observing that the quotient space 
${\hat AdS}_{2,n}=AdS_{2,n}/Z_n$ which is isomorphic to $AdS_2$ and has codimension two orbifold singularities, $E_n$ which are pointlike[\REV,\DOB]. These singularities can be created by codimesion two cosmic zero branes which are located at $E_n$ and have tension $T_n=(n-1)/4G_2n$. The presence of the branes modifies the geometry of ${\hat AdS}_{2,n}$. If we can compute this metric then
we can obtain the modular entropy given by[\REV]
$$S^n_{mod}=\partial_n I_{EH}({\hat AdS}_{2,n}) \quad, \eqno(21)$$
and use the relation $S_{ent}=S^n_{mod}(n=1)$.

The metric near the zero branes is (in Euclidean polar coordinates)
$$ds^2=r^2dt^2+ n^2 dr^2 \quad. \eqno(22)$$
The curvature diverges at the brane unless the extrinsic curvature vanishes which means $E_n$ is an extremal surface as required by the holographic entanglement entropy. Since the metric is singular at the location of the zero branes, we consider an infinitesimal surface, $E_n(\epsilon)$, (which is a circle of radius $\epsilon$) around them that acts as a UV cutoff. Now, since the space--time has a boundary we have to add to the Einstein--Hilbert action the Gibbons--Hawking boundary term on $E_n(\epsilon)$[\EUC]
$$I_{bndy}={1 \over {8 \pi G_2}} \int_{E_n(\epsilon)} dx~ K \quad, \eqno(23)$$
where $K$ is the trace of the extrinsic curvature with $K=1/n \epsilon$ on the cutoff circle. Then we find
$$I_{bndy}(E_n(\epsilon))=I_{bndy}({\hat AdS}_{2,n})={1 \over {8 \pi G_2}}{{2 \pi \epsilon} \over {n \epsilon}}={1 \over {4 G_2n}}
\quad. \eqno(24)$$
Using[\REV,\DOB] and $I_{grav}=I_{EH}-I_{bndy}$
$$S^n_{mod}=\partial_n I_{grav}({\hat AdS}_{2,n})=-\partial_nI_{bndy}({\hat AdS}_{2,n})={1 \over {4G_2n^2}} \quad, \eqno(25)$$
we find $S_{ent}=1/4G_2=A_H/4G_D$ as required.

We showed that the asymptotic infinity of \S black holes are Weyl equivalent to $AdS_2 \times S^{D-2}$ and these two 
space--times have the same thermodynamics. Then, we found that the entanglement entropy
of $AdS_2$ exactly reproduces the black hole entropy. We stress that, in this description, the black hole degrees of freedom live at asymptotic infinity and not near or on the horizon in contrast to every other description of 
\S black hole entropy. Of course, it is well--known that in the AdS/CFT correspondence, black hole
entropy counts the degrees of freedom on the boundary, i.e. at asymptotic infinity. It is gratifying to see that a similar picture arises for \S black holes in flat space--time. Our result crucially depends on the fact that \S black holes are asymptotically Weyl equivalent to 
$AdS_2 \times S^{D-2}$ and not higher dimensional $AdS$ space--times since these do not carry any entanglement entropy (due to the fact that they have only one connected boundary).

It is important to note that \S black holes are asymptotically flat and only Weyl equivalent to $AdS_2 \times S^{D-2}$. 
On the other hand, we saw that the Weyl transformation in eq. (4) does not change the thermodynamics, i.e. the temperature and entropy as a function of the black hole mass. Thus, it is reasonable to conclude that the \S black hole entropy can be described by that of $AdS_2 \times S^{D-2}$. In fact, 
the entanglement entropy we computed above holds for a whole class of space--times that are related by Weyl transformations. \S black holes and asymptotically $AdS_2 \times S^{D-2}$ are simply two examples related by eq. (4) in this class. 
In general. the metrics in this class are related by Weyl transformations that
only asymptotically (i.e. for $r \to \infty$ in the original coordinates) reduce to eq. (4). This is the minimal requirement for leaving the entropy invariant. In general, these Weyl transformations lead to complicated metrics that describe fibrations; however, asymptotically they all 
reduce to $AdS_2 \times S^{D-2}$ with the correct radii. Eq. (4) is the simplest Weyl transformation in this class.
This situation is perhaps similar to critical phenomena in which there are universality classes of
distinct microscopic theories that are described by the same low energy physics. Critical phenomena can be completely described 
by only a few macroscopic parameters, i.e. the critical exponents without
any knowledge about the microscopic physics. 

The only property of the \S black hole metric that is crucial for our result is the fact that it is asymptotically flat, i.e.
$f(r)=1+g(r)$ where $g(r) \to 0$ as $r \to \infty$. This property guarantees that under the Weyl transformation in eq. (4) 
and $r \to r_0^2/r$ we obtain a metric that is asymptotically (as $r \to 0$) $AdS_2 \times S^{D-2}$. As a result, 
we can apply the same method to count the entropy of all nonextremal, asymptotically flat black objects. 
Clearly, the method does not apply to
black holes in anti de Sitter and de Sitter spaces since these are not asymptotically flat.

We obtained the entropy of \S black holes with nonzero temperature from that of $AdS_2 \times S^{D-2}$ which has vanishing temperature. Surprisingly, the origin of black hole entropy seems not to be thermal but rather the nonvanishing entanglement in the $AdS_2$ vacuum. 
This is a result of the asymptotic limit $r \to 0$ we took above. Eq. (5), after dimensional reduction over $S^{D-2}$, describes an
$AdS_2$ black hole at finite temperature. On the boundary, this corresponds to an excited (or thermal) state. The asymptotic limit
$r \to 0$, is the IR limit in the bulk in which all the information about the $AdS_2$ black hole is lost. As we saw above, this is 
the $r \to \infty$ limit of the global $AdS_2$. On the boundary, this corresponds to the UV limit in which all thermal effects are negligible and only the high energy modes of the vacuum are relevant. These give rise to the entanglement entropy in eq. (20) which is finite in $AdS_2$. Thus, the fact that \S black hole entropy arises due to the entanglement in the vacuum is a direct result of the asymptotic limit we took in order to describe the entropy by degrees of freedom at infinity.

Even though we found that the entropy of \S black holes is the entanglement entropy of $AdS_2$, we do not have a clear idea about the degrees of freedom that we count. The machinery of holographic entanglement entropy allows us to compute the entropy of $AdS_2$ 
without any knowledge about the degrees of freedom that live on its boundary.
Following the AdS/CFT correspondence, we expect that the boundary of $AdS_2$ is described by a one dimensional CFT with only a time coordinate, i.e. conformal quantum mechanics[\CQM] which is not well--understood. However, by compactifying the much better understood $AdS_3$ to $AdS_2$ (on its boundary $S^1$), this theory was shown to be equivalent to a chiral or light--cone 2D CFT[\CHI]. It seems that one can also count the $AdS_2$ entropy in certain situations in string theory[\SEN]. Clearly, the nature of the one--dimensional boundary theory dual to $AdS_2$ and the degrees of freedom its entanglement entropy counts are very important questions that require further investigation.

It is interesting to compare our results with the more conventional derivation of \S black hole entropy as entanglement entropy which has a rich literature[\SOL]. First, the conventional entanglement entropy describes the entanglement between degrees of freedom just inside and outside the horizon. In our case, the entangled degrees of freedom reside on the two boundaries of an $AdS_2$ that lives at asymptotic infinity.
Second, the conventional entanglement entropy is a UV cutoff dependent quantity that diverges in the continuum limit. 
It is a one--loop correction to the black hole entropy[\REN] and reproduces it completely only in the context of induced gravity[\JAC].
In our case, the holographic entanglement entropy of $AdS_2$ naturally introduces a cutoff and gives a finite result that is the correct entropy. 
Third, conventional entanglement entropy depends on the type and number of quantum fields that are assumed to live near the horizon which leads to the species problem. In our case, we do not need to know the degrees of freedom on the $AdS_2$ boundary to compute the holographic entanglement entropy and there is no species problem.

\bigskip
\centerline{\bf 3. Holographic Entanglement Entropy in Generalized Theories of Gravity}
\medskip

We now show that the method used in the previous section also applies to black holes in generalized theories of gravity with higher derivative terms in the action. In these theories, the black hole entropy is given by the Wald entropy[\WAL]
$$S_{Wald}=-2 \pi \int_H \sqrt{h} {{\partial {\cal L}} \over {\partial R_{abcd}}} \epsilon_{ab} \epsilon_{cd}  \quad, \eqno(26)$$
where the integral is over the black hole horizon $H$ with the metric $h_{\mu \nu}$ and 
$\epsilon_{\mu \nu}= \zeta_{\mu} \eta_{\nu}-\zeta_{\nu} \eta_{\mu}$ where $\zeta_{\mu}$ is the Killing vector of the horizon and 
$\eta_{\nu}$ is its normal such that $\zeta_{\mu} \eta^{\mu}=1$.

In order show that we can obtain Wald entropy using our method we first assume that at 
asymptotic infinity, these black holes, just like \S black holes, are Weyl equivalent to $AdS_2 \times S^{D-2}$.
On general grounds, for a black hole in a generalized theory of gravity, $f(r)= 1+ g(r^{-a})$ where $g(r^{-a})$ vanishes asymptotically for $r \to \infty$. This is simply the requirement of asymptotic flatness of the black hole metric in these theories. It is easy to see that after the Weyl transformation in eq. (4)
and $r \to r_0^2/r$, we find that asymptotically (i.e. in the limit $r \to 0$) $f(r) \to r^2/r_0^2$ leading to the $AdS_2 \times S^{D-2}$ space--time.

As an example, we consider 
Gauss--Bonnet gravity with the Lagrangian[\GB]
$${\cal L}=-{1 \over {16 \pi G}} [R + \alpha(R_{abcd} R^{abcd}-4R_{ab}R^{ab}+R^2)] \quad, \eqno(27)$$
where $\alpha$ is the coefficient of the Gauss--Bonnet term.
Gauss--Bonnet black hole solutions exist for $D \geq 5$ with the metric[\GB]
$$ds^2=-f(r)~dt^2+f(r)^{-1} dr^2+ r^2 h_{ij} dx^idx^j \quad. \eqno(28)$$
The indices $i,j$ run over the $D-2$ transverse directions with a metric of constant curvature $h_{ij}$ equal to $(D-2)(D-3)$. Black hole solutions are defined by
$$f(r)=1+{r^2\over {2 \alpha}} \left(1-\sqrt{1+{{64 \pi G \alpha M} \over {(D-2)V_k r^{n-1}}}} \right)  \quad, \eqno(29)$$
where $V_k$ is the unit volume in k dimensions and with an abuse of notation $\alpha$ has been rescaled by a factor of $(D-3)(D-4)$.

The radius of the black hole horizon, $r_0$, is determined by the largest value that satisfies $f(r_0)=0$. 
The Wald entropy of Gauss--Bonnet black holes has been calculated in ref. [\GB] to be
$$S={{r_0^{D-2} V_k} \over {4G}} \left(1+{{2 \alpha (D-2)} \over {(D-4)r_0^2}}\right) \quad. \eqno(30)$$

Now, looking at the form of $f(r)$ in eq. (29) it is easy to see that after the Weyl transformation in eq. (4) and the coordinate
transformation $r \to r_0^2/r$, in the limit $r \to 0$ we again find $f(r) \to r^2$ leading to an $AdS_2 \times S^{D-2}$
space--time. This means that, just like for \S black holes, we can compute the entropy of Gauss--Bonnet black holes from the entanglement entropy of $AdS_2$ .

We now repeat the procedure of the previous section and calculate the entropy using eqs. (17) and (19).
On the n--sheeted $AdS_2$, the Riemann tensor becomes[\TAK]
$$R_{abcd}=R^{0}_{abcd}+2 \pi (1-n)(g_{ac} g_{bd}- g_{ad} g_{bc}) \delta_H \quad, \eqno(31)$$
where $R^0$ is the Riemann tensor of ordinary $AdS_2$ including the contribution from the cosmological constant and $\delta_H$
is a delta function over the horizon. Using the relation $g_{ab}=\zeta_a \eta_b+\zeta_b \eta_a$ one can show that
$\epsilon_{ab} \epsilon _{cd}=-(g_{ac} g_{bd}-g_{ad} g_{bc})$. If we expand the gravitational Lagrangian ${\cal L}(R_{abcd})$ in powers of 
$R_{abcd}$ around $R^0_{abcd}$, the terms higher than the linear one do not contribute to the entanglement entropy since
${d \over dn}(1-n)^m$ vanishes in the limit $n \to 1$ for $m \geq 2$. As a result, the gravitational action becomes
$$I_{grav}= 2 \pi (1-n) \int_H \sqrt{h} {{\partial {\cal L}} \over {\partial R_{abcd}}} \epsilon_{ab} \epsilon_{cd}  \quad, 
\eqno(32)$$
leading to the entanglement entropy
$$S_{ent}=-{\partial \over \partial n} (-I_{grav})=S_{Wald} \quad. \eqno(33)$$

Thus, in generalized theories of gravity, black holes with asymptotically flat metrics are also asymptotically Weyl equivalent to 
$AdS_2 \times S^d$. The entanglement entropy of the $AdS_2$ factor exactly reproduces the Wald entropy of the black hole. As metioned above, this is not true for black holes that are not asymptotically flat such as those in de Sitter and anti de Sitter space--times.

In theories beyond General Relativity, black hole entropy is not proportional to horizon area and the notion of one bit of information per Planck area clearly fails in these cases. Therefore, the generalization of holography to higher derivative theories does not seem to be straightforward. However, we see that black hole entropy continues to be given by the holographic entanglement entropy of $AdS_2$. Perhaps this is the notion of holography that is common to all theories of gravity.

The more precise method of computing $S_{ent}$ by the modified $AdS_2$ metric due to cosmic zero branes can also be generalized to theories of gravity beyond General Relativity[\ORD]. One again uses cosmic D0 branes in the bulk of $AdS_2$ but now with generalized equations of gravity and obtains the Wald entropy. Since this is quite technical and we have nothing new to add we refer the reader to the original literature[\REV,\ORD].

\bigskip
\centerline{\bf 4. The Holographic Entanglement Entropy and Other Methods} 
{\bf of Entropy Computation}
\medskip

It is interesting to note that the computations done in section 2, even though conceptually very different, are technically quite similar to the calculations of black hole entropy using the methods of conical deficit angle[\CON] and Euclidean gravity[\EUC]. In Euclidean gravity, the near horizon region of a \S black hole, which is Rindler space, is described by the flat metric in polar coordinates with Euclidean time parametrized by the angle. In this context, the angular period is the inverse of the Hawking temperature, $1/T_H$ with the period for flat space set to be $2 \pi$ .
In order to do thermodynamics, we should be able to vary the temperature and therefore the periodicity of Euclidean time from $2 \pi$
[\CON]. Thus, we take the angular periodicity to be $2 \pi \alpha$ where the deficit angle is $2 \pi (1-\alpha)$. 
The Einstein--Hilbert action in the near horizon region (reduced over the transverse sphere which is the horizon) is given by
$$I_{EH}=-{1 \over {16 \pi G_2}} \int_{R_{\alpha}} d^2x (R+ \Lambda) \quad, \eqno(34)$$
Here $\Lambda=0$ since Rindler space is flat and $R_{\alpha}$ is the Rindler (flat) space with a conical deficit angle of
$2 \pi (1-\alpha)$. $G_2=G_D/A_H$ is the
two--dimensional Newton constant obtained by the dimensional reduction over the horizon area $A_H$.
Now using the change in the Riemann scalar due to the conical deficit
$$R \to R + 4 \pi (1 - \alpha) \delta^2(x) \quad, \eqno(35)$$
we find
$$\int_{R_{\alpha}} d^2x R= \alpha \int_R d^2x R + 4 \pi (1-\alpha)  \quad, \eqno(36)$$
where $R$ is the Rindler (flat) space without the conical deficit angle.
The gravitational action becomes
$$I_{EH}=-{1 \over {4 G_2}} (1 - \alpha) \quad, \eqno(37)$$
Since in Euclidean gravity the action represents (the negative of) the free energy[\EUC], the entropy is given by
$$S=-(\alpha \partial_{\alpha}-1) (-I_{EH})={1 \over {4 G_2}}={A_H \over {4G_D}} \quad, \eqno(38)$$
which is exactly the black hole entropy.

The similarities between the method of conical deficit angle and the one used in section 2 are now clear. In fact, eqs. (18) and (30) are virtually identical due to the fact that the Riemann scalar is modified the same way in the presence of conical singularities. Therefore, the gravitational actions which are identified with the free energies are the same in both cases leading to the same entropy. However, the origins
of the deficit angles are different in each case. When we compute the entanglement entropy of a boundary of $AdS_2$, we use the replica trick under which we consider the n--sheeted $AdS_2$. This introduces a cut with the deficit angle $2 \pi (1-n)$.
In the conical deficit angle method, the conical deficit
of $2 \pi (1-\alpha)$ arises due to the variation of the temperature in the near horizon geometry which is Rindler space. These expressions are formally identical leading to identical entropies.

On the other hand, the two methods are conceptually very different. The method in section 2 counts degrees of freedom at 
(Weyl transformed) asymptotic infinity whereas the conical deficit method counts those near the horizon. We do not know what these degrees of freedom are in either case. The entangled degrees of freedom of $AdS_2$ live on its boundaries and may be described by a conformal quantum mechanics or a chiral 2D CFT. The ones counted by the conical deficit method are even more obscure since this is a purely formal manipulation of the Euclidean metric without any reference to degrees of freedom.

The second and more precise calculation of $S_{ent}$ using the modified geometry due to the cosmic zero branes is also technically very reminiscent of the calculation of black hole entropy in Euclidean gravity. First, we note that, in both the calculation in section 2 and and in Euclidean gravity, black hole entropy arisies solely from the boundary term in the gravitational action. In
Euclidean gravity, black hole entropy is given by[\EUC]
$$S_{BH}={1 \over {8 \pi G_D}} \int_H d^Dx \sqrt{-h}~ K = {1 \over {8 \pi G_D}} {\partial_r} \int_H d^Dx \sqrt{-h}              \quad, \eqno(39)$$
where the integral is over the horizon and $h$ is the horizon metric. The second equation above compared to eq. (25) shows the similarity between the two calculations. Of course, in Euclidean gravity the derivative is with respect to the normal direction to the
horizon (i.e. the radial direction) whereas in our calculation the derivative is with respect to $n$ which is the number of sheets
of $AdS_2$ (generalized to be continuous).
Near the horizon, the Rindler metric in dimensionless units, i.e. with the surface gravity $\kappa=1$, is given by
$$ds^2= r^2 d\tau^2+ dr^2 +r_H^2 d\Omega^2 \quad, \eqno(40)$$
which is again very similar to the metric near the zero branes in eq. (22). Using eq. (39) and the fact that the integral along the Euclidean time and transverse directions give $2 \pi$ and $A_H$ respectively we find as expected the correct black hole entropy.
Thus the two computations use very similar boundary actions, definitions of entropy and metrics and lead to the same entropy.
However, they are conceptually very different. $S_{ent}$ is the entanglement entropy which is computed by a holographic method. 
$S_{BH}$ computed in Euclidean gravity has no relation to holography (other than the result). The geometries used in the two methods are also different. In Euclidean gravity, the geometry is Euclidean Rindler space with the horizon at the origin. In the holographic entanglement method, the geometry is the n--sheeted $AdS_2$ where the origin is the location of the zero brane.

\bigskip
\centerline{\bf 5. Conclusions and Discussion}
\medskip

In this paper, we showed that under a Weyl and coordinate transformation, the \S metric becomes asymptotically $AdS_2 \times S^{D-2}$.
We assumed that the thermodynamics of \S black holes and the asymptotically $AdS_2 \times S^{D-2}$ space--times are the same since they have the same temperature and entropy. Then, we found that the entanglement entropy of $AdS_2$ ($\times S^{D-2}$) precisely reproduces the \S black hole entropy. It is surprising that the origin of black hole entropy seems not to be due to any thermal effects but due to the entanglement in the ($AdS_2$) vacuum. Generalization of our results to black holes in generalized theories of gravity seems to indicate that the holographic principle is still valid but only in the sense of the holography on the asymptotic $AdS_2$ and not on the black hole horizon.

We stress that the results of this paper crucially depend on the assumption that we are allowed to Weyl transform the \S metric so that the thermodynamics of \S black holes can be described by the asymptotic $AdS_2 \times S^{D-2}$. This is supported by the fact that if we dimensionally reduce 4D General Relativity to two dimensions we get 2D dilatonic gravity which has an $AdS_2$ black hole
solution which is given exactly by eq. (5) (plus a dilaton profile)[\CAD]. In this case, one Weyl transforms the 2D black hole metric by the same factor in eq. (4). Thus, it seems reasonable to use the Weyl and coordinate transformations above to investigate the \S black hole. We note that the results ref. [\CAD] hold only for 4D \S black holes and it would be interesting to find out whether they can be generalized to higher dimensions.  

Our method can be generalized to any asymptotically flat metric in a straightforward manner. It is
easy to show that any such metric will become, in the asymptotic limit, $AdS_2 \times S^{D-2}$ under the combined Weyl and coordinate transformations employed above. Therefore, the entropies of all nonextremal black objects with asymptotically flat metrics can be obtained from the entanglement entropy of $AdS_2$[\LAS]. Notably, this is not the case for black holes in de Sitter and anti de Sitter space--times since these are not asymptotically flat. $AdS$ black holes are of course asymptotically $AdS$ but (except for $AdS_2$ black holes) these are higher dimensional $AdS$ space--times with one connected boundary and vanishing entanglement entropy. 

It is well--known that the near horizon geometries of extremal black holes contain $AdS_2$ components which seem to be the origin of their entropies[\NHO,\DAB]. In this paper, we showed that the entropies of \S black holes are also given by the the entanglement entropy of $AdS_2$. Since, as noted above,  our results can be generalized to all nonextremal black holes, it seems that the entanglement entropy of $AdS_2$ constitutes the universal origin the entropy of all black holes (with asymptotically flat metrics).  

In our description \S black hole entropy arises from the entanglement of the two boundaries of $AdS_2$.
In fact, the full space--time is $AdS_2 \times S^{D-2}$ if we include the $S^{D-2}$ that we integrated out at low energies. This makes it clear that the entanglement is actually between the two $S^{D-2}$s at either boundary of $AdS_2$. Each Planck area of a sphere at one boundary is maximally entangled with a corresponding Planck area of the sphere at the other boundary. This picture is very similar to the one we have for the conventional entanglement entropy for black holes but in our case the spheres are not just inside and outside the horizon but at asymptotic infinity. The holographic result of one bit per Planck area is simply due to the form of the entanglement entropy of $AdS_2$, i.e. $S_{ent}=1/4G_2$. As a result, we should perhaps think of the geometry as an $AdS_2$ stretched between two $S^{D-2}$s. In the spirit of the ER=EPR idea[\EPR], the $AdS_2$ is the geometric representation of the entanglement between the two spheres.

It is strange to find that the entropy of \S black holes are due to degrees of freedom that live at infinity. One way to understand this is to think of a spherical shell that starts to collapse from infinity to form the black hole. If we assume that, at the beginning of the collapse the shell is in a pure state but entangled with degrees of freedom at infinity, after the black hole is formed we are left with a mixed state since the degrees of freedom of the shell are behind the horizon and have to be traced over. As a result, the entropy of the black hole is completely due to the entanglement of the shell which is behind the horizon and
the degrees of freedom at infinity. In such a scenario, it would not be surprising to find that the entanglement entropy at infinity gives exactly the correct entropy for \S black holes.

It has been shown that the near horizon region with the geometry of Rindler space ($\times S^{D-2}$) can be described by a CFT that  reproduces \S black hole entropy[\LAST]. The results of this paper provide an alternative description of black hole entropy in terms of degrees of freedom that live at infinity. These two descriptions differ by the replacement of Rindler space by $AdS_2$. 
Interestingly, both of these spaces are described by two dimensional CFTs but with different central charges.
Moreover, Rindler space is related to a specific $AdS_2$ black hole (Rindler $AdS_2$ space) by a coordinate and Weyl transformation[\LAST] which is similar to the relation between \S black hole metrics and asymptotically $AdS_2 \times S^{D-2}$ examined in this paper. Thus, we find that the near horizon physics is described by an $AdS_2$ black hole whereas
that at asymptotic infinity is described by global $AdS_2$. This leads to the intriguing possibility that the near horizon and the asymptotic descriptions are the IR and UV limits of the same CFT that are related by a renormalization flow. 
Another way to reach the same conclusion is to note that the near horizon region of 4D \S black holes can be transformed into $AdS_2 \times S^2$ by a Kinnersley transformation[\KIN]. Here the $AdS_2$ part is actually Rindler $AdS_2$ space whereas the spherical factor is the horizon just as in our case. Thus, Kinnersley transformations lead us to exactly the same situation described above.


\bigskip
\centerline{\bf Acknowledgments}

I would like to thank the Stanford Institute for Theoretical Physics for hospitality.

\vfill

\refout

\end
\bye